\documentclass[manuscript]{aastex62}
\graphicspath{{./}{figures/}}

\received{July 18, 2018}
\revised{September 7, 2018}

\submitjournal{ApJ}

\shorttitle{Distance to 47 Tuc and NGC 362}
\shortauthors{Chen et al.}

\begin{document}

\title{Distances to the globular clusters 47 Tucanae and NGC 362 using Gaia DR2 parallaxes}

\correspondingauthor{Seery Chen}
\email{seery.chen@mail.mcgill.ca}

\author{Seery Chen} 
\affiliation{Department of Physics and Astronomy, University of British Columbia, Vancouver, BC V6T 1Z1}

\author{Harvey Richer}
\affiliation{Department of Physics and Astronomy, University of British Columbia, Vancouver, BC V6T 1Z1}

\author{Ilaria Caiazzo}
\affiliation{Department of Physics and Astronomy, University of British Columbia, Vancouver, BC V6T 1Z1}

\author{Jeremy Heyl}
\affiliation{Department of Physics and Astronomy, University of British Columbia, Vancouver, BC V6T 1Z1}



\begin{abstract}
Using parallaxes from Gaia DR2, we estimate the distance to the globular clusters 47~Tuc and NGC~362, taking advantage of the background stars in the Small Magellanic Cloud and quasars to account for various parallax systematics. We found the parallax to be dependent on the Gaia DR2 G-band apparent magnitude for stars with $13<G<18$, where brighter stars have a lower parallax zero point than fainter stars. The distance to 47~Tuc was found to be  $4.45\pm0.01\pm0.12$~kpc, and for NGC 362 $8.54\pm0.20\pm0.44$~kpc with random and systematic errors listed respectively. This is the first time a precise distance measurement directly using parallaxes has been determined for either of these two globular clusters.

\end{abstract}



\section{Introduction} \label{sec:intro}

47 Tucanae (47 Tuc) and NGC 362 are two globular clusters seen projected in front of the Small Magellanic Cloud (SMC). Previously, the distance to 47 Tuc has been estimated by methods such as main sequence fitting \citep{Percival}, white dwarf spectral energy distributions \citep{Woodley}, RR Lyrae stars \citep{Bono}, eclipsing binaries \citep{Thompson}, and various other techniques \citep[see][for a summary of previous methods]{Woodley}. For NGC 362, the distance has been estimated using RR Lyrae stars \citep{Szekely}. However, a precise distance measurement directly using parallaxes for either of these clusters has never been obtained, due to a required parallax precision of tens of microarcseconds.

Gaia Data Release 2  \citep[Gaia DR2,][]{gaiacollab,gaiacollab2016} provided the five astrometric parameters (position, parallax, and proper motion) for more than 1.3 billion sources.  In particular the parallaxes, with a median uncertainty  0.04 milliarcseconds (mas) for bright ($G<14$~mag) sources, can be used to determine distances to an unprecedented number of objects.  However, for more distant and fainter stars, the parallaxes become sensitive to systematic errors. These systematics are significant when determining the distance to 47~Tuc and NGC~362.

Globally, the parallax zero point was found by \citet{Lindegren} to be $-0.029$~mas, in the sense that Gaia parallaxes are too small. However, adding a global zero point to the data is insufficient as the zero point depends on the position on the sky. It can vary by as much as 0.1~mas globally and  0.04~mas on intermediate ($<20\deg$) scales and small ($<1\deg$) scales \citep{X. Luri}. \citet{Lindegren} also found a possible dependence on colour and magnitude which can cause variations of $0.02$~mas.

Fainter objects tend to have a much larger uncertainty in measured parallax. Since quasars tend to be fainter, this makes using them to account for all of these spatial parallax systematics difficult. As there is only a small number of quasars behind 47~Tuc and NGC~362, quasars are insufficient to account for small-scale parallax zero point variations; thus, we employ the SMC stars behind the clusters to account for these systematics. 


The basic premise of this paper is to find the distance to 47~Tucance and NGC~362 by using quasars to account for the intermediate scale parallax systematics, and the SMC stars behind each cluster to account for the small-scale parallax systematics, and to further investigate the colour and magnitude dependent parallax systematics to obtain a precise distance estimate to 47~Tuc and NGC~362 using trigonometric parallax. 

\section{Data}\label{sec:Data Selection}
\subsection{Selecting SMC and Quasars}

For the SMC to quasar comparison, a circular field of 5 degrees in radius was taken around the SMC. Quasars were identified from a cross match with the ALLWISE catalog \citep{Secrest} found on the Gaia archive (\texttt{gaiadr2.allwise\_best\_neighbour}) \citep{gaiacollab}. SMC stars were chosen by a proper motion selection; furthermore, only stars with G-band apparent magnitude (G-mag) brighter than 19 were used, as fainter stars have a much larger parallax spread. Finally, a $5\sigma$ parallax error cut was applied to both SMC stars and quasars, where $\sigma$ is the standard deviation of the parallax distribution.

\begin{figure}[ht!]
	\plottwo{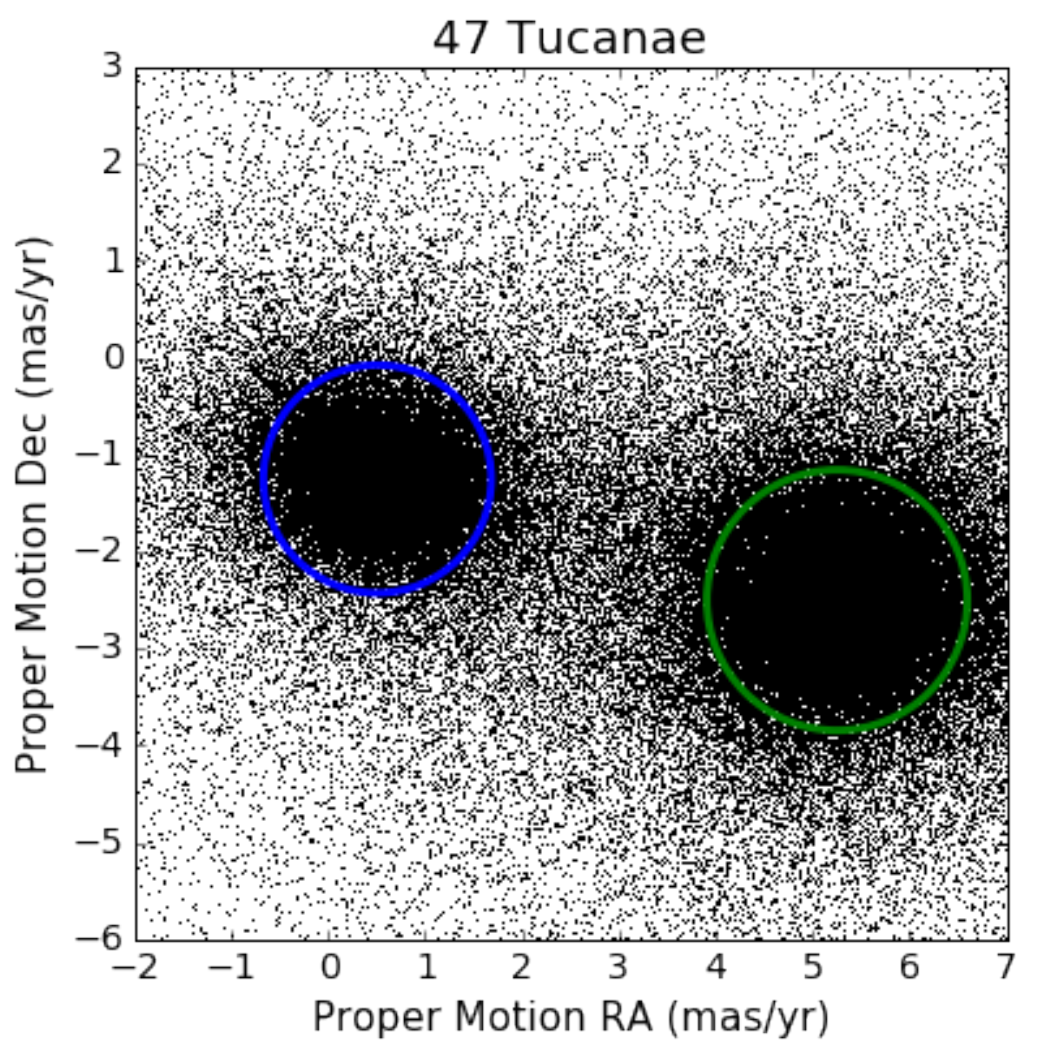}{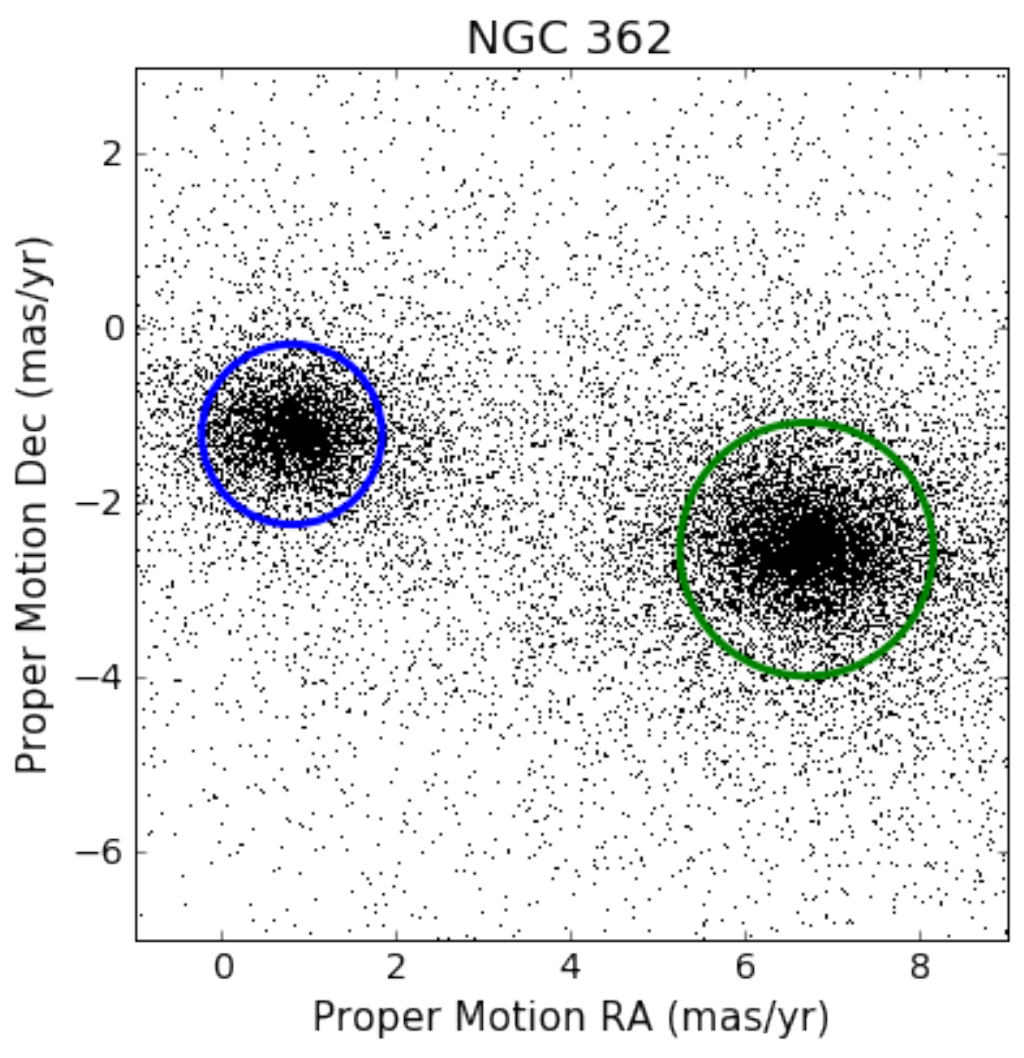}
	\caption{Proper motion selections for the SMC (shown in blue) and foreground cluster stars (in green). The centre of each proper motion circle was found by fitting two Lorentzian peaks in proper motion in RA and Dec, and the radius was taken to be twice the peak's half maximum width. See Appendix A, figure \ref{fig:peakfit} for peak fit.}
	\label{fig:pm}
\end{figure}

\begin{figure}[ht!]
	\plottwo{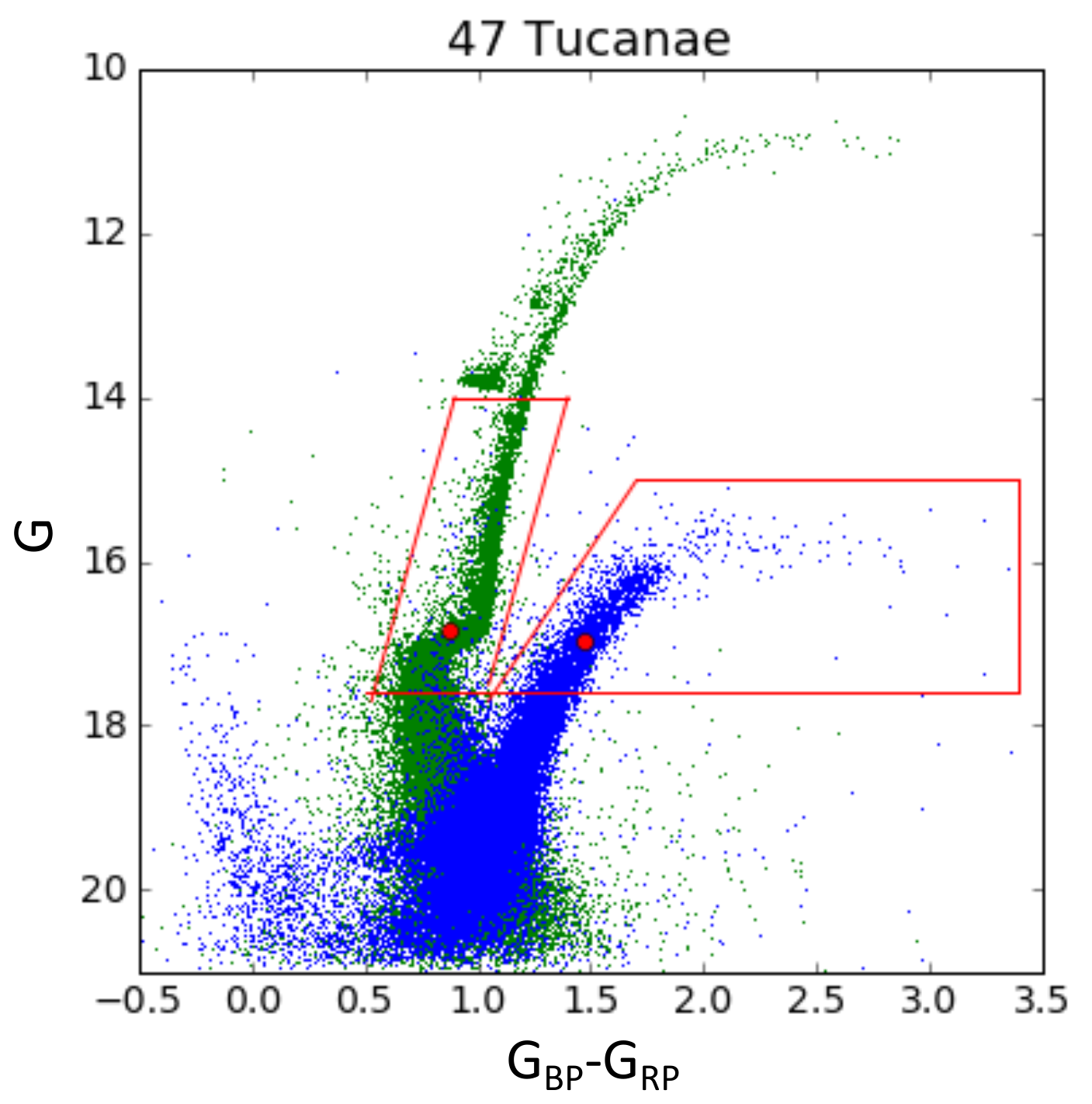}{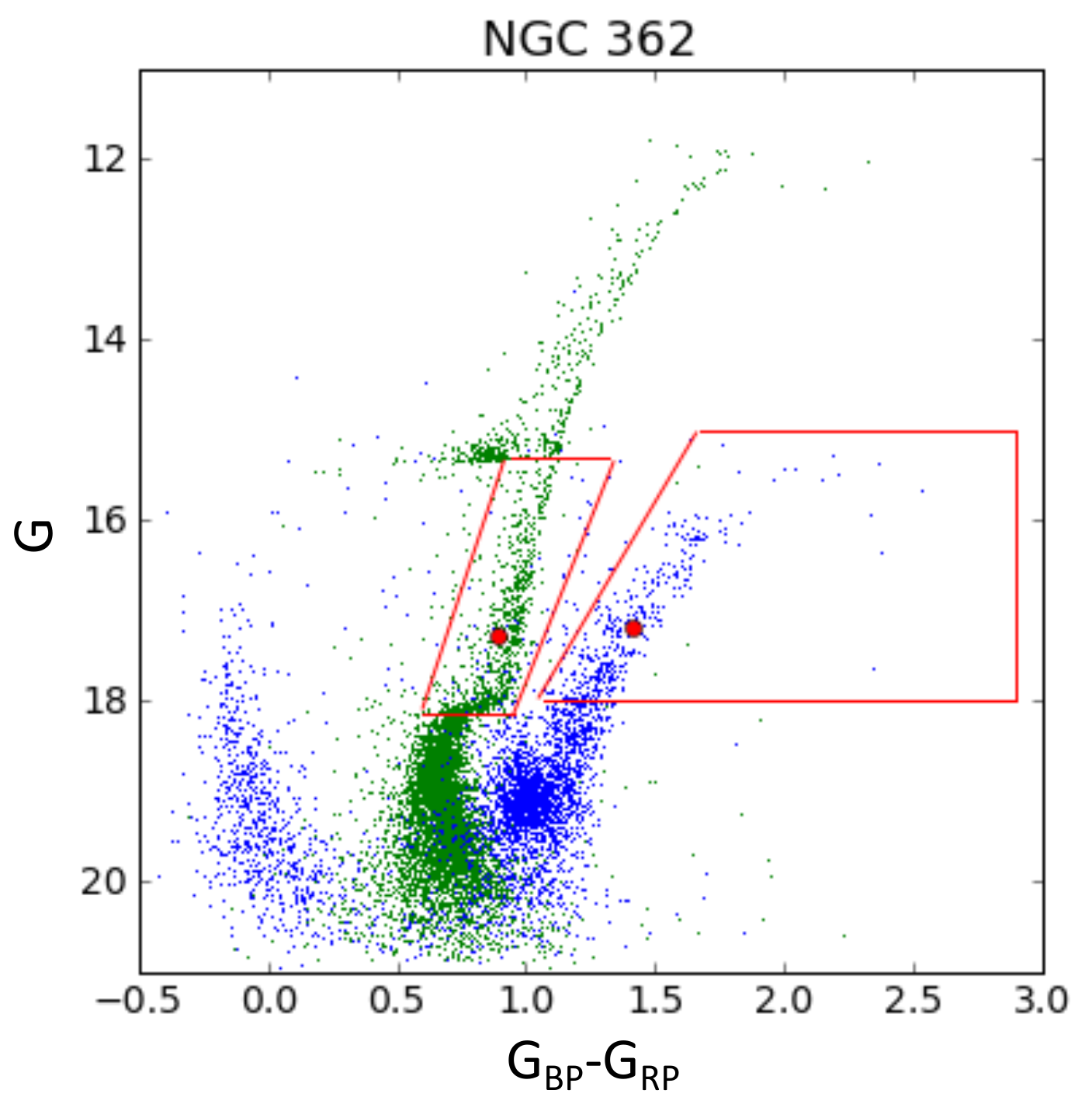}
	\caption{Colour magnitude diagrams for cluster and SMC stars after the proper motion selection was applied.  47 Tuc and NGC 362 are in green, and the SMC is in blue. The selected stars with the same mean G-mag used in the subsequent analysis is shown by the red boxes. The red dots show the mean Gaia DR2 $G_{BP}-G_{RP}$ colour and G-mag of each selection.}
	\label{fig:cmd}
\end{figure}

\subsection{Selecting 47 Tucane, NGC 362 and SMC Stars}

For 47 Tuc, NGC 362, and the SMC stars behind each cluster, the following cuts were applied to obtain the selections used for the analysis. First, for 47~Tuc, stars within one degree of the centre of the cluster were selected. For NGC 362, only stars within 0.3~degrees of the centre were selected. Second, for both clusters, a proper motion cut was applied to separate cluster stars from SMC stars (see figure~\ref{fig:pm}). To avoid magnitude dependent systematics, stellar selections were chosen to have the same mean G-band apparent magnitude (see figure \ref{fig:cmd}). Finally a $3\sigma$ cut in parallax was applied to remove outliers in each sample.


\section{Colour and Magnitude Systematics}

\citet{Lindegren} found that the parallax zero point appeared to vary depending on colour and magnitude. We chose our selection of cluster stars and SMC stars to have the same magnitude to avoid this possible systematic. However, as these selections do not have the same average colour, we further investigate the possible zero point dependence on colour and magnitude. 

\begin{figure}[ht!]
\includegraphics[width=\textwidth]{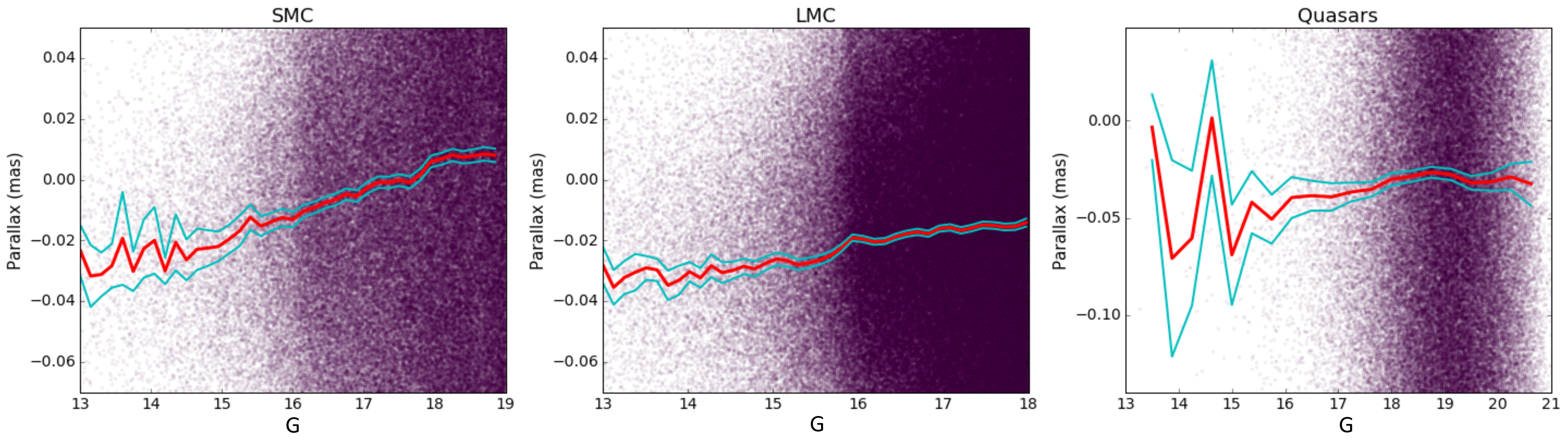}
\caption{ The parallax as a function of G-band apparent magnitude for the five-degree selection of SMC stars, LMC stars, and all quasars over the entire sky matched to the ALLWISE catalog. The running mean is shown in red, and $2\sigma$ uncertainty shown in cyan. For $13<G<19$, the running mean was fit to a line, with slopes from left to right respectively being $0.00779\pm0.00027$, $0.00399\pm0.00015$, and $0.00545\pm0.00051$ mas/G-mag.}
	\label{fig:g_run}
\end{figure}


In figure \ref{fig:g_run}, we present a plot of parallax vs G-band apparent magnitude for quasars in the ALLWISE catalog \citep[c.f.][]{Lindegren}, and did the same for the SMC and the Large Magellanic Cloud (LMC). The LMC data were selected in the same manner as the SMC data in section \ref{sec:Data Selection}. All three plots show the same trend of brighter stars having a lower parallax zero point. 

The linear trend for G-mag vs parallax only appears for stars with $G<18$. The average magnitude of our five-degree SMC selection used to determine the SMC parallax ($\pi_{smc}$) is 17.9 in G-mag, and the quasars are even fainter. Thus we concluded that the magnitude dependent systematic in calculating $\pi_{smc}$ is insignificant. As we chose our selection of 47~Tuc stars and NGC 362 stars to have the same average G-mag as the SMC selection behind each cluster, the magnitude dependence does not affect our results. See appendix B for further discussion of the magnitude-parallax systematic for selections that do not have the same G-mag. 

\begin{figure}[ht!]
\includegraphics[scale = 0.35]{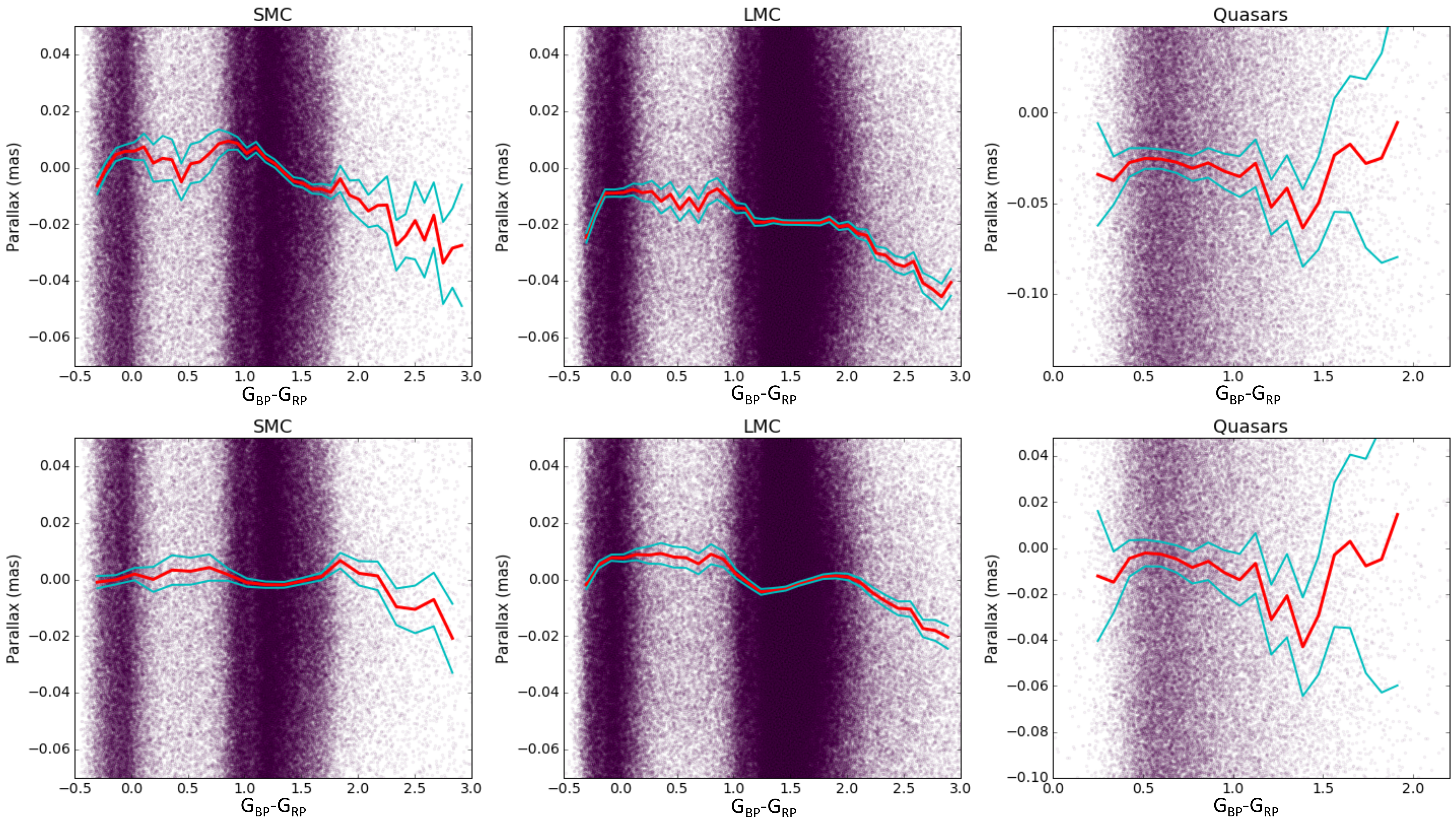}
\caption{ Parallax as a function of $G_{BP}-G_{RP}$ colour for the five-degree selection of SMC stars, LMC stars, and all quasars matched to the ALLWISE catalog. The running mean is shown in red, and $2\sigma$ uncertainty shown in cyan. Top row is without the magnitude correction. The bottom row is with a magnitude correction intended to set the parallax to zero.}
	\label{fig:bprp_run}
\end{figure}

When plotting parallax vs $G_{BP}-G_{RP}$ colour, there initially appeared to be a trend for the SMC (see figure \ref{fig:bprp_run}). This can partially be explained by the red giant branch of the SMC where stars tend to get redder as they get brighter. When the parallax dependence on magnitude was accounted for, the section between 1 and 1.5 in colour no longer has a downward trend. When applying this correction to the LMC and quasar selection, it is unclear as to whether applying the magnitude correction eliminates the possibility of the parallax being dependent on colour. However these effects appear to be minimal between 0.5 and 1.5 in $G_{BP}-G_{RP}$ colour, thus we do not account for it in our final result. 

\section{Analysis}\label{sec:Analysis}

Let $\pi_{smc}$, $\pi_{47}$ and $\pi_{362}$ be, respectively, the true SMC, 47~Tuc and NGC~362 parallaxes. Since quasars should have a parallax of essentially zero, we subtract the weighted mean quasar parallax from the weighted mean SMC parallax, using $\frac{1}{\sigma_{\pi}^2}$ as the weight, where $\sigma_{\pi}$ is the error in parallax given by the 5 parameter astrometric fit. This accounts for a spatially dependent parallax zero point systematic error across the SMC, giving us $\pi_{smc}$. The error in averaging the SMC parallax incorporates any systematic errors arising from the depth of the SMC.

\begin{figure}[ht!]
	\plottwo{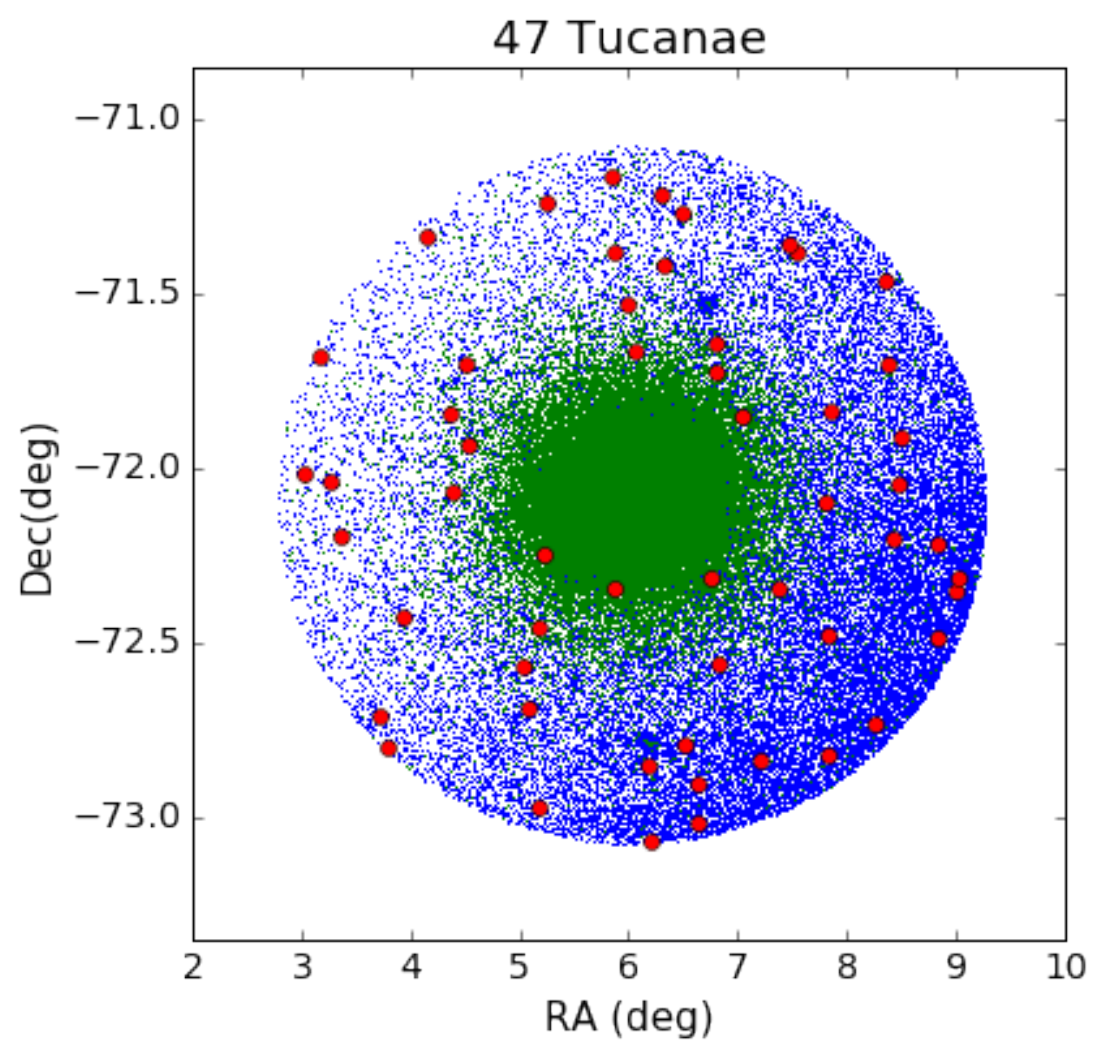}{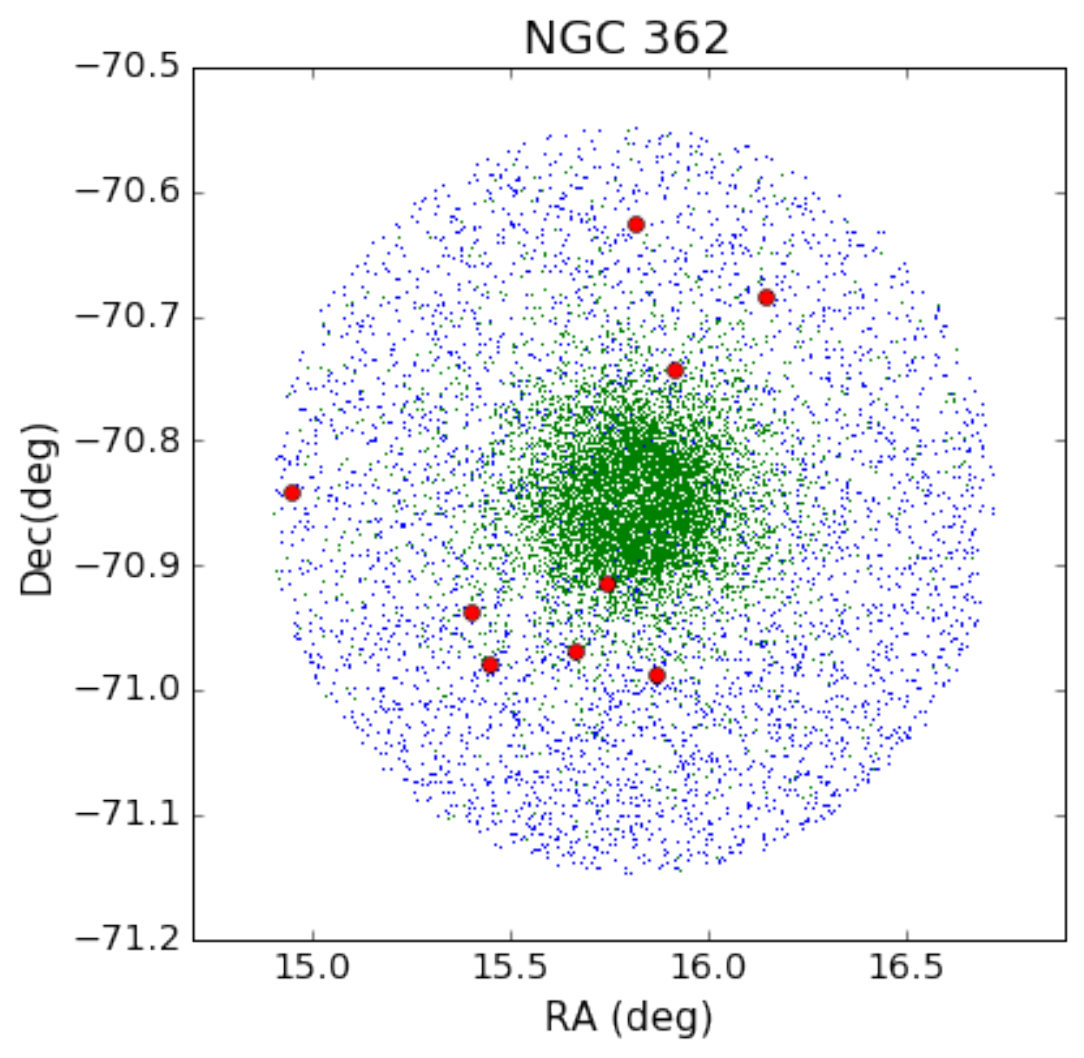}
	\caption{SMC stars (in blue) and foreground cluster stars (in green) within 1 degree of 47 Tuc on the left, and within 0.3 degrees of NGC 362 on the right. The red circles show the quasars behind each selection.}
	\label{fig:ra_dec}
\end{figure}

Assuming uniform small-scale variations across the one-degree 47 Tuc selection, the weighted mean over our SMC selection can be subtracted from the weighted mean over our 47 Tuc selection to get $\pi_{47}-\pi_{smc}$.

However, as the distribution of SMC stars behind 47 Tuc is non-uniform (see figure \ref{fig:ra_dec}), this could introduce systematic errors.
To account for non uniform small scale parallax zero point variations we use a pairwise method. Let $\hat{\pi} ^i_{47}$ represent the measured parallax for the $i$th star in the 47 Tuc selection, then
\begin{equation}
\hat{\pi}_{47}^i=\pi_{47}+\delta_{int}+\delta_{small} 
\end{equation}
where $\delta_{int}$ is the intermediate scale parallax zero point offset, and $\delta_{small}$ is the small scale spatially dependent parallax offset. Pairing up each 47 Tuc star with the nearest SMC stars, we eliminate the intermediate and small scale parallax variations, as stars close in RA and Dec should have the same small and intermediate scale parallax zero point offset. 
The  $n$ nearest SMC stars to each 47 Tuc star have parallaxes of $\hat{\pi}_{smc}^j$ with respective errors $\sigma_{smc}^j$. The weighted mean of those $n$ stars is $\hat{\pi}_{smc}^i$, such that
\begin{equation}
\hat{\pi}_{smc}^i=\frac{\sum_{j=1}^{n} w_j \hat{\pi}_{smc}^j}{\sum_{j=1}^{n} w_j} 
= \pi_{smc}+\delta_{int}+\delta_{small} 
\end{equation}
where $w_j=\left(\frac{1}{\sigma_{smc}^i}\right)^2$ is the weight, and $\sigma_{smc}^i=\sqrt[]{\frac{1}{\sum_{j=1}^{n} w_j}}$ is the error in the weighted mean. The $n$ SMC stars should all have the same $\delta_{int}$ and $\delta_{small}$. Subtracting the $i$th mean SMC parallax from the $i$th 47 Tuc parallax, and adding their respective errors in quadrature to get the random error of $\sigma_i$, we take the weighted mean over $N$ 47~Tuc stars with $w_i=\frac{1}{\sigma_i^2}$ to get 
\begin{equation}
\frac{\sum_{i=1}^{N} w_i (\hat{\pi}_{47}^i-\hat{\pi}_{smc}^i)}{\sum_{i=1}^{N} w_i} =\pi_{47}-\pi_{smc}.
\end{equation}
Adding the value of $\pi_{smc}$ found previously, we find $\pi_{47}$ and thus the distance to 47 Tuc.

One issue with the pairwise method is that it could double-count SMC stars. Another method to account for a small scale parallax zero point is to divide the selection into squares and subtract the weighted mean of SMC and 47 Tuc parallaxes in each square, then take a weighted mean over all the squares to get $\pi_{47}-\pi_{smc}$.

Applying the above three different methods to NGC 362 stars instead of 47 Tuc stars gives us estimates for $\pi_{362}$, and the distance to NGC 362. 

\section{Results}\label{sec:Results}

\subsection{SMC and Quasars}
The weighted average of the SMC parallax was found to be $-0.0059 \pm 0.0001$ mas and for the quasars was $-0.0251\pm0.0060$ mas. The difference gives $\pi_{smc}=0.0192 \pm 0.0060$ mas which corresponds to a distance of $52 ^{+23}_{-12}$ kpc. 

\subsection{47 Tucane}

In the pairwise analysis, we used a search radius of 0.1 deg around each 47 Tuc star, where the mean parallax of all SMC stars within 0.1 deg of each 47 Tuc star was subtracted from the parallax of that 47 Tuc star. Using search radii from 0.02 to 0.1 deg all gave results which agreed within $3\sigma$ error (see appendix A, figure \ref{fig:pairwise}).

For the third method of dividing the selection into squares and subtracting the average SMC and 47 Tuc parallaxes in each square, we used 16 squares of 0.15 deg on a side. The following results shown in Table~\ref{tab:47summary}, and numbers quoted in the remainder of the paper, are listed with random and systematic errors listed respectively (where the systematic errors result from the uncertainties in the parallax of the SMC and the zero points).
\begin{table}[h!]
\renewcommand{\thetable}{\arabic{table}}
\centering
\caption{Summary of results for 47 Tucanae.}
\label{tab:47summary}
\begin{tabular}{cCCC}
\tablewidth{0pt}
\hline
\hline
Method &  $\pi_{47}-\pi_{smc}$  (mas) & $\pi_{47}$ (mas) &  $d_{47} (kpc)$ \\ 
\hline
\decimals
Weighted Mean    & 0.2070\pm0.0013 & 0.2262\pm0.0013\pm0.0060 & 4.42\pm0.02\pm0.12 \\ 
Pairwise         & 0.2055\pm0.0006 & 0.2247\pm0.0006\pm0.0060 & 4.45\pm0.01\pm0.12 \\
Squares          & 0.2075\pm0.0022 & 0.2267\pm0.0022\pm0.0060 & 4.31\pm0.04\pm0.12 \\ 
\hline
\end{tabular}
\end{table}

\begin{table}[ht!]
\renewcommand{\thetable}{\arabic{table}}
\centering
\caption{Summary of results for NGC 362.}
\label{tab:362summary}
\begin{tabular}{cCCC}
\tablewidth{0pt}
\hline
\hline
Method &  $\pi_{362}-\pi_{smc}$  (mas) & $\pi_{362}$ (mas) &  $d_{362} (kpc)$ \\ 
\hline
\decimals
Weighted Mean    & $0.0988\pm0.0046$ & $0.1178\pm0.0046\pm0.0060$  & $8.49\pm0.33\pm0.43$  \\ 
Pairwise         & $0.0981\pm0.0028$ & $0.1171\pm0.0028\pm0.0060$  & $8.54\pm0.20\pm0.44$  \\
Squares          & $0.1001\pm0.0049$ & $0.1191\pm0.0049\pm0.0060$  & $8.39\pm0.35\pm0.42$  \\

\hline
\end{tabular}
\end{table}

\subsection{NGC 362}
Repeating the processes used for 47 Tuc on NGC 362, we derive the results shown in Table \ref{tab:362summary}. For the pairwise method we again used a search radius of 0.1 degree around each NCG 362 star. For squares, we divided the sample into 16 squares of 0.1 deg on a side since the sample was much smaller.

\section{Discussion}\label{sec:Discussion}

\subsection{SMC parallax}

An SMC parallax of $\pi_{smc}=0.0192 \pm 0.006$ mas corresponding to a distance of $52 ^{+23} _{-12}$ kpc agrees within $1\sigma$ with the distance estimate of $62.1 \pm 1.9$ kpc given by \citet{Graczyk}. The large uncertainty in our estimate is primarily due to the parallax spread of the quasars. While our result agrees with the literature values, it is not a particularly insightful result and serves primarily as a way to continue onto the distance determination of 47 Tuc and NGC 362 using parallax measurements. Using the SMC distance from \citet{Graczyk} to calculate the distances to 47~Tuc and NGC~362 gives distances of $4.51 \pm 0.02$ kpc and $8.76 \pm 0.22$ kpc respectively, with random and systematic errors combined. This does shift our values to slightly further distances, however they are still well within $1\sigma$ of our model independent values determined directly with parallax. 

\subsection{Comparison with Literature Values}
Our result for 47 Tuc, $4.45\pm0.01\pm0.12 $ kpc, is close to average for 47 Tuc distance estimates, which range between $4.29\pm0.47$~kpc (estimated  kinematically by \citet{Heyl}) and $4.94\pm0.25$~kpc\footnote{distance in kpc converted from distance modulus, full compiled list of 47 Tuc distance modulus found in \citet{Woodley}} found by \citet{Bono} using RR Lyrae stars. The most precise literature value comes from horizontal branch fitting from near-IR photometry, where \citet{Salaris2007} found a distance of $4.33\pm0.06\pm0.05$~kpc\textcolor{xlinkcolor}{$^1$} with random and systematic errors listed respectively. While this distance yields a somewhat smaller error estimate than our result, it is model dependent whereas our distance estimate is not. See Appendix A, figure \ref{fig:compare} for a comparison of our distances with those in the literature from the past 20 years.

For NGC 362, our distance of $8.54\pm0.20\pm0.44$ kpc agrees within $1\sigma$ of the literature values -  \citet{Harris} (2010 edition) quotes 8.6 kpc (no error reported) and $7.9\pm0.6$ kpc \citet{Szekely} using RR Lyrae stars.

Our distance to 47~Tuc is smaller than the RR Lyrae distance, while for NGC~362 it is larger when compared with the RR Lyrae distance from the literature. This could be the result of these authors using two different techniques. \citet{Bono} used K band photometry of a single RR Lyrae star in 47 Tuc while \citet{Szekely} used V band photometry for multiple RR Lyrae in NGC 362.  

\subsection{Cluster Properties}

From our parallax measurements, the difference in distance moduli between NGC 362 and 47 Tuc is $1.415\pm 0.048$. From the CMDs, using Gaia photometry, we get a mean difference of $1.446\pm 0.004$ in magnitude between the red horizontal branch stars of NGC 362 and 47 Tuc.  These values agree within 1$\sigma$ and thus we cannot see significant metallicity effects on the red horizontal stars branch from this comparison.  In any case, our distance uncertainties are too large to probe the modest differences in the magnitudes of RR~Lyrae stars expected from theoretical models \citep{Marconi} over the metaliticity range spanned by 47~Tuc and NGC~362.

Our distances will have a direct impact on calculating the absolute cluster age. Our somewhat larger distance modulus for NGC 362 compared with the RR Lyrae distance from \citet{Szekely}, for example, suggests a more luminous turnoff magnitude and hence a slightly younger age for the cluster.  


\section{Conclusions}\label{sec:Conclusions}
In deriving distances to the globular clusters 47 Tuc and NGC 362, we needed to account for the spatial and magnitude dependent parallax systematics in Gaia DR2. To accomplish this we did the following three things: 
\begin{enumerate}
\item Took a weighted mean of the quasars behind the SMC which allowed us to find an intermediate scale parallax zero point of $-0.0251\pm0.0060$ mas, yielding $\pi_{smc}=0.0192\pm0.0060$ mas
\item Took a selection of foreground cluster stars with the same mean G mag as the selection of SMC stars to avoid magnitude dependent parallax systematics
\item Paired up all SMC stars within 0.1 degrees of each cluster star to account for the small scale parallax zero point variations
\end{enumerate}
The parallax zero point was not significantly dependent on colour for stars between 0.5 and 1.5 in $G_{BP}-G_{RP}$ colour which the majority of stars in our selections lie between. This is not to say that the parallax zero point is entirely independent of colour, further investigation into a possible colour dependence would be needed for such a statement. 

This yields the distance estimate of $4.45\pm0.01\pm0.12$ kpc for 47 Tuc and $8.54\pm0.20\pm0.44$~kpc for NGC 362, with random and systematic errors quoted respectively. This is currently the most precise distance determination to NGC 362 available. While our  estimate for 47 Tuc is not more precise than some previous estimates, it is comparable in precision and not model dependent.


\acknowledgments


This work has made use of data from the European Space Agency (ESA) mission
{\it Gaia} (\url{https://www.cosmos.esa.int/gaia}), processed by the {\it Gaia}
Data Processing and Analysis Consortium (DPAC,
\url{https://www.cosmos.esa.int/web/gaia/dpac/consortium}). Funding for the DPAC
has been provided by national institutions, in particular the institutions
participating in the {\it Gaia} Multilateral Agreement.

This project was developed in part at the 2018 Gaia Sprint, hosted by the eScience and DIRAC Institutes at the University of Washington, Seattle.

We acknowledge the support of the Natural Sciences and Engineering Research Council of Canada (NSERC).

%





\newpage

\appendix

\section{Additional Figures}

\begin{figure}[ht!]
	\plottwo{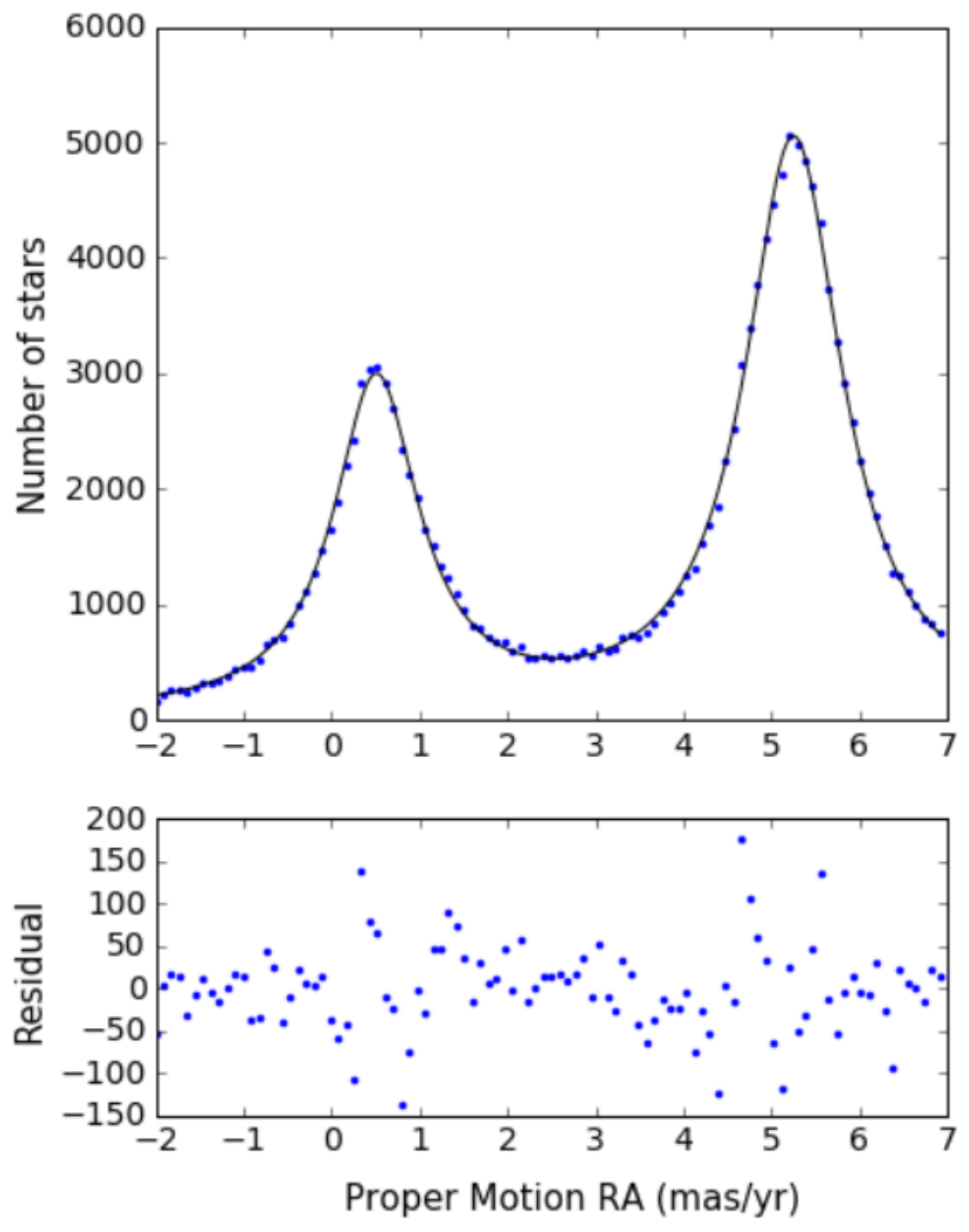}{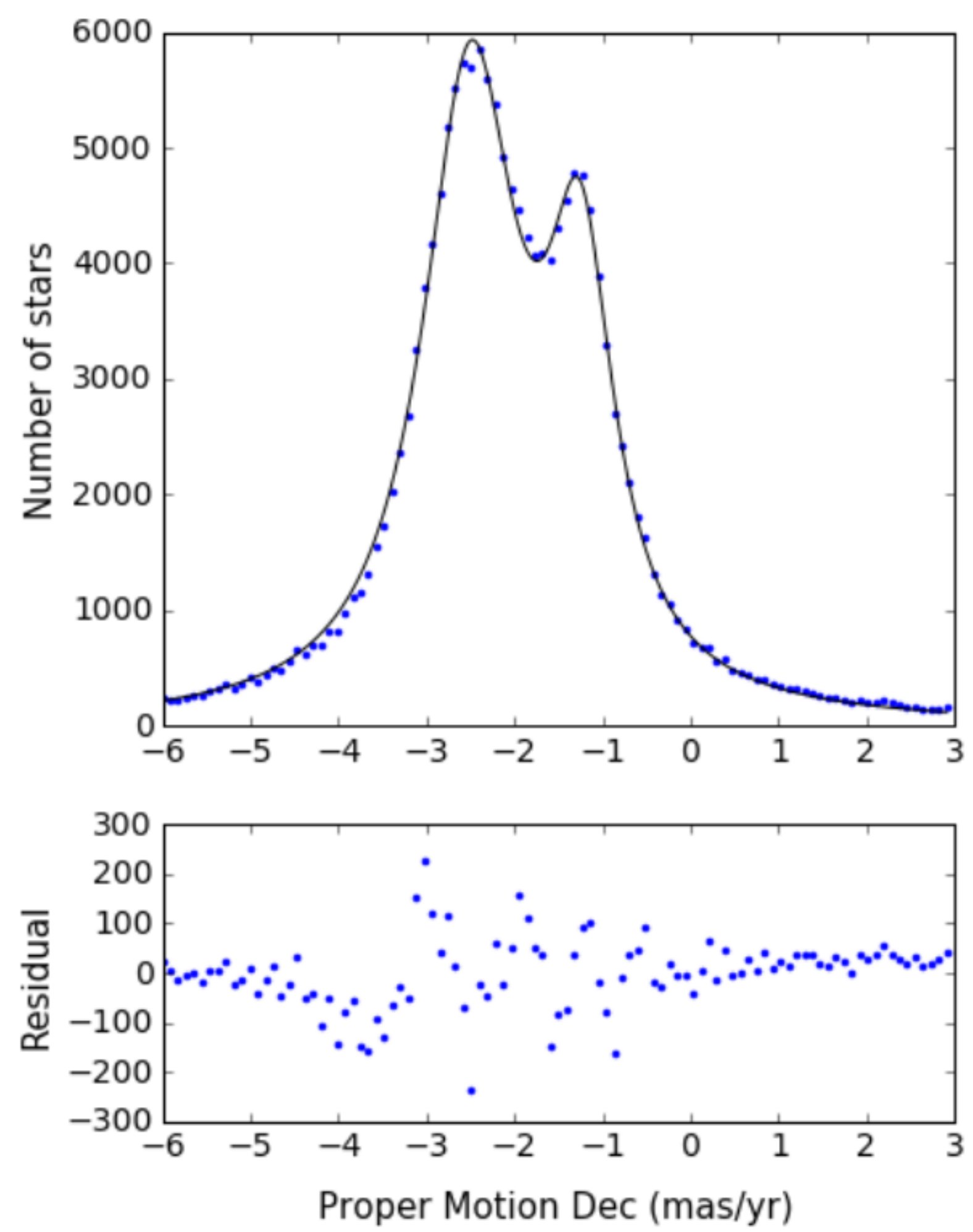}
	\caption{Lorentzian peak fit of proper motion right ascension for all stars within 1 degree of 47 Tuc on the left, and in proper motion declination on the right.}
	\label{fig:peakfit}
\end{figure}

\begin{figure}[ht!]
	\plottwo{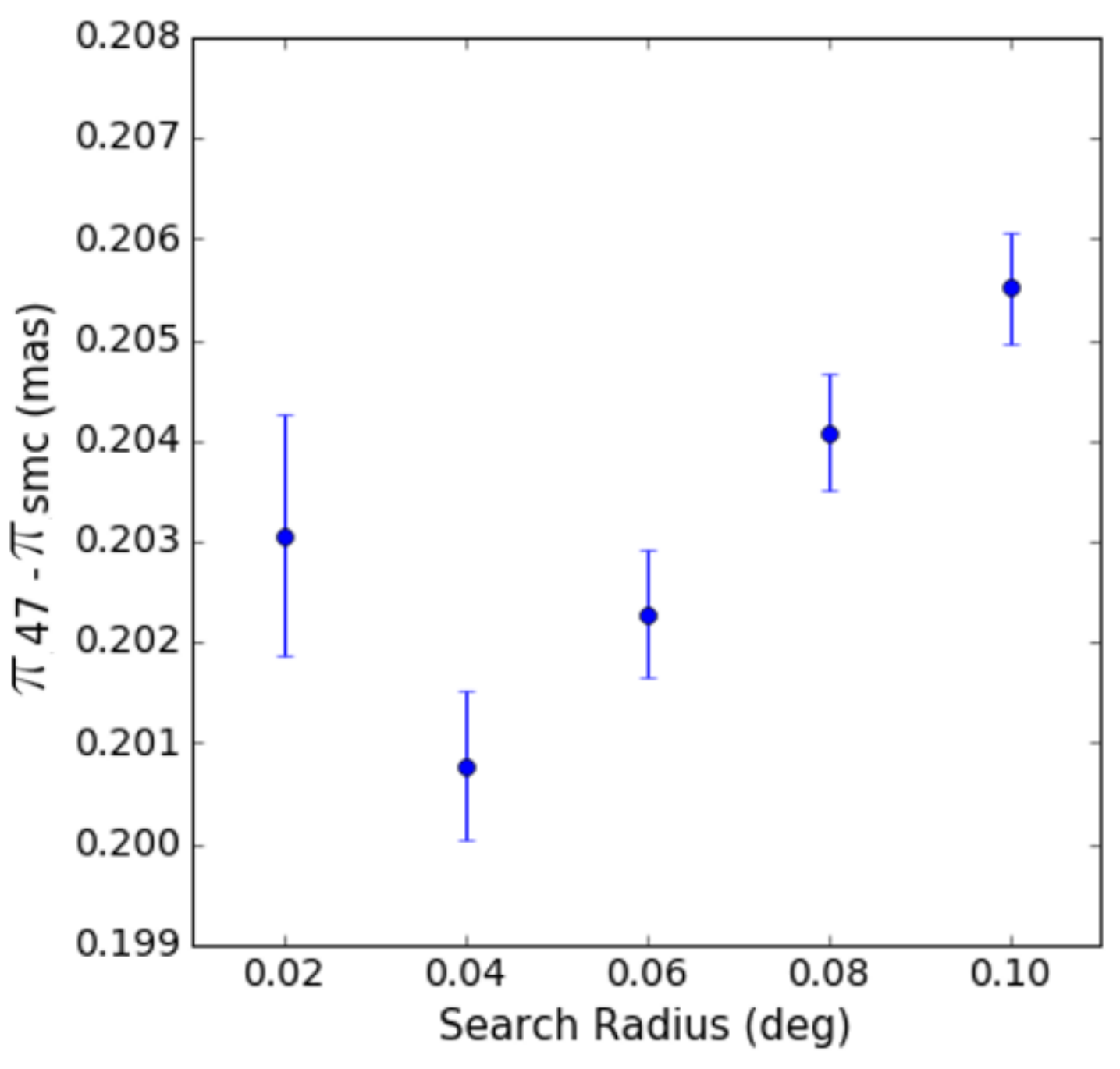}{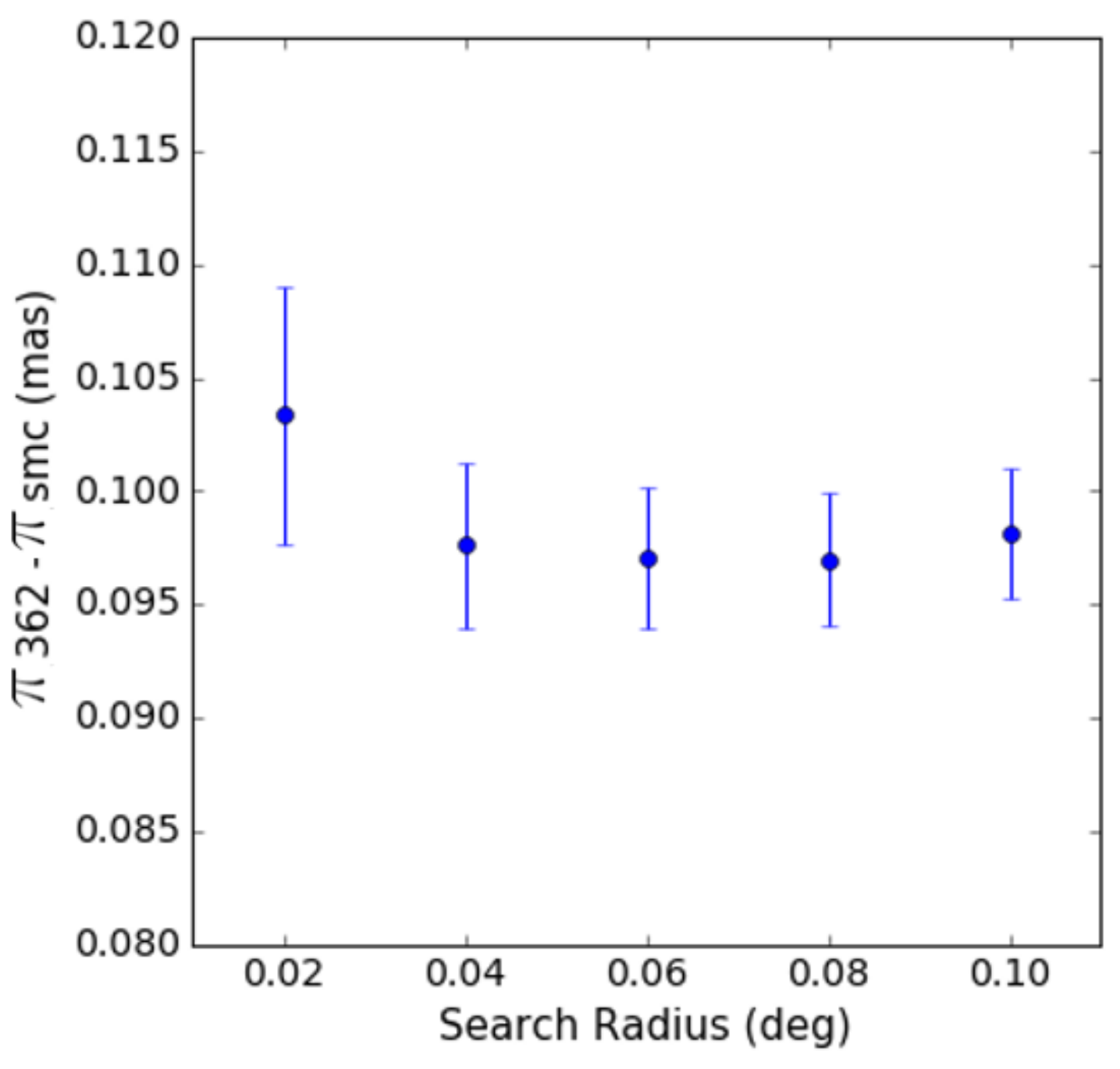}
	\caption{Results for $\pi_{47}-\pi_{smc}$ and $\pi_{362}-\pi_{smc}$ using varying search radii in pairwise analysis. Error bars show the $1\sigma$ uncertainty.}
	\label{fig:pairwise}
\end{figure}

\begin{figure}[ht!]
	\plottwo{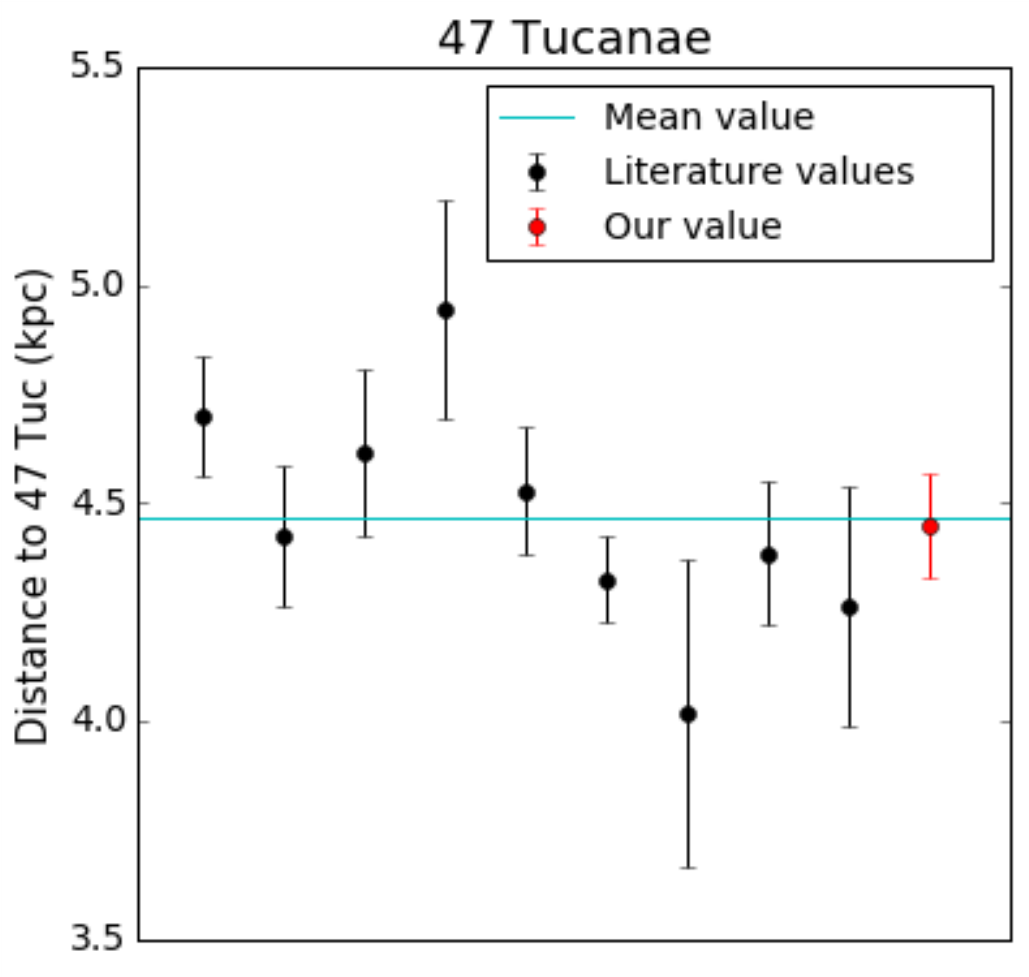}{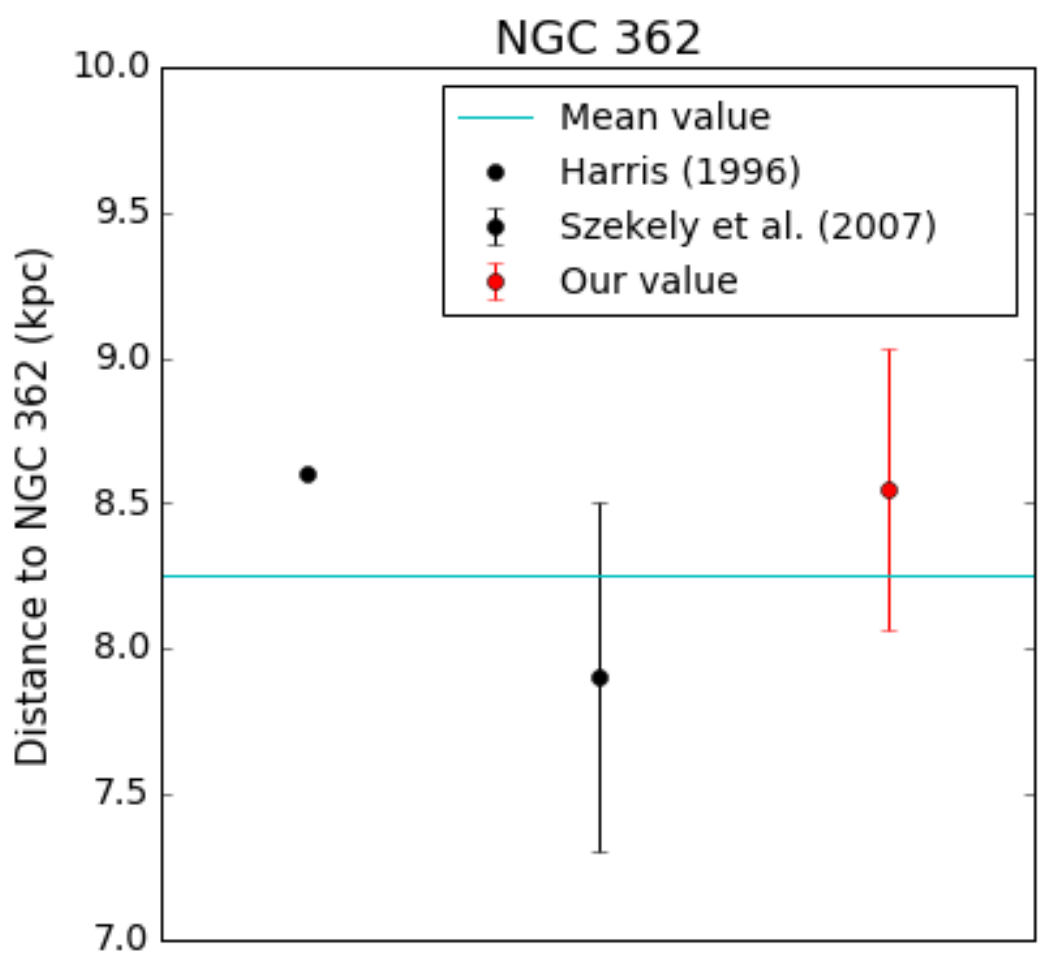}
	\caption{Literature values for distances to 47 Tuc and NGC 362 from the past 20 years. Distances for 47~Tuc converted from distance modulus, found in table 1 of \citet{Woodley}, where values cited from left to right are: \citet{Zoccali}, \citet{Grundahl}, \citet{McLaughlin}, \citet{Salaris2007}, \citet{Kaluzny}, \citet{Bono}, \citet{Thompson}, \citet{Woodley}.}
	\label{fig:compare}
\end{figure}

\newpage

\section{Method to account for magnitude dependent parallax systematics}

\begin{figure}[ht!]
	\plottwo{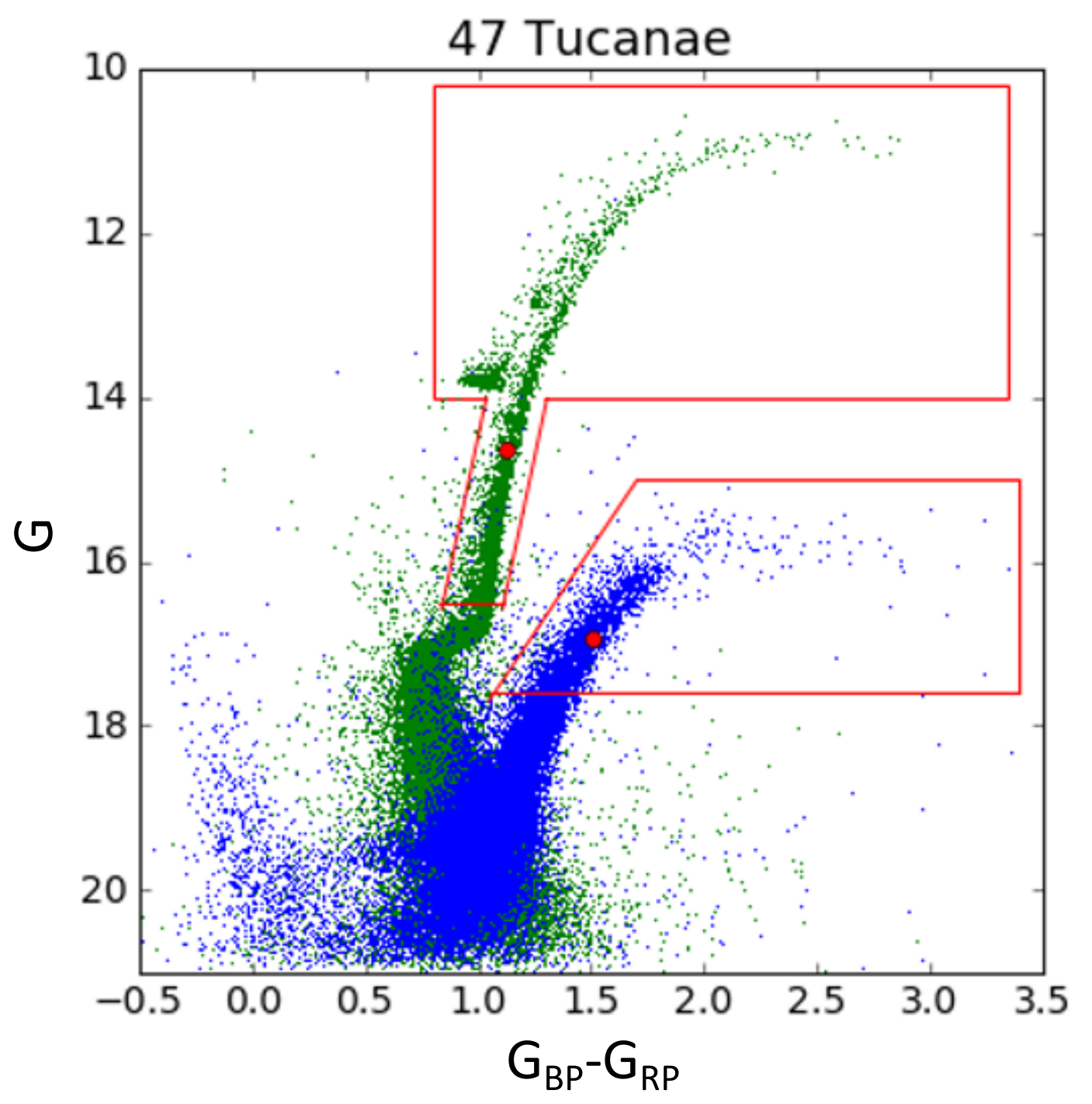}{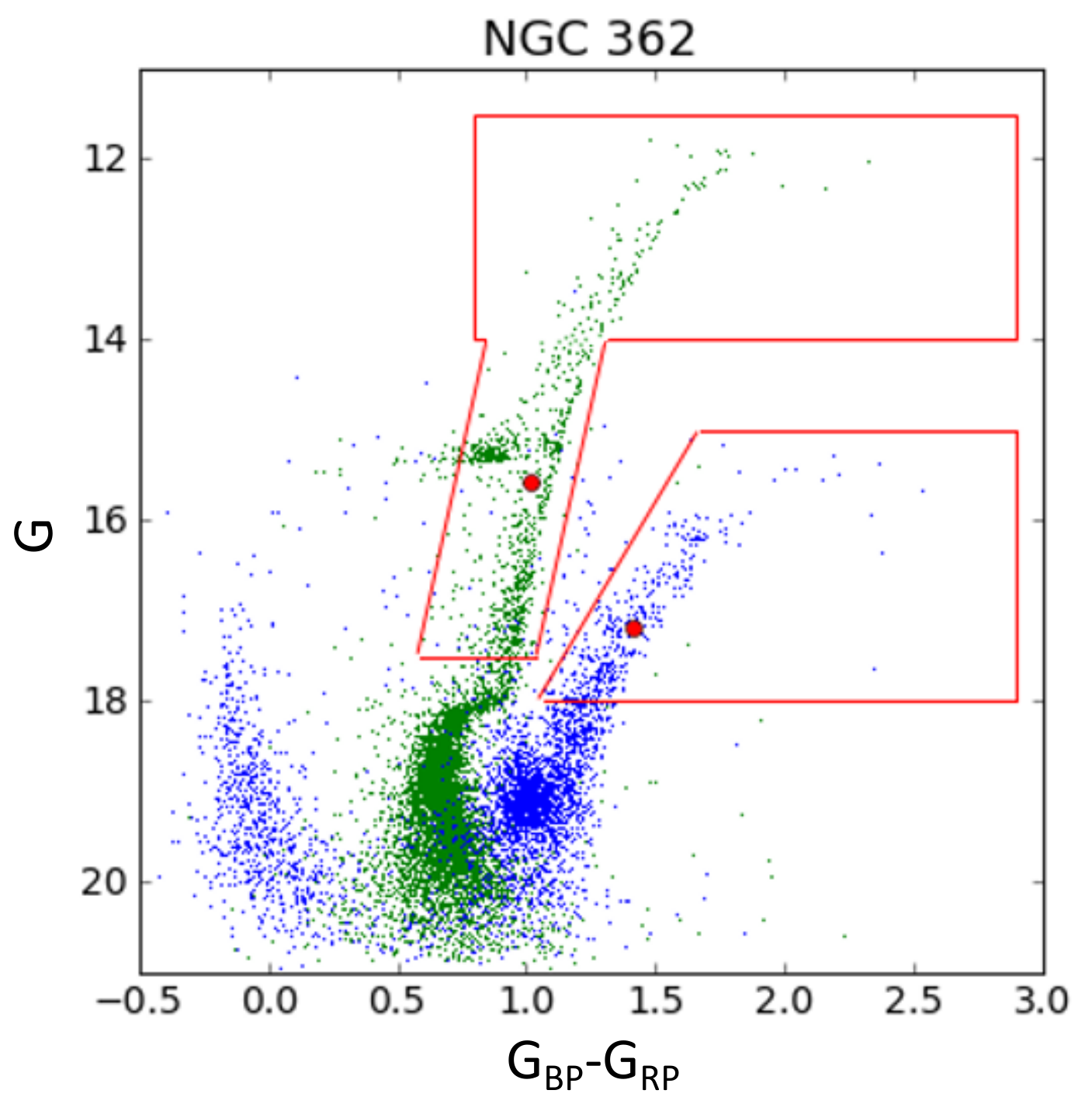}
	\caption{Colour magnitude diagram for cluster and SMC stars after the proper motion selection was applied.  47 Tuc and NGC 362 are in green, and the SMC is in blue. The original selection of stars with different G mag are shown by the red boxes. The average G~mag and $G_{BP}-G_{RP}$ colour are shown by the red circles. }
	\label{fig:47cmd2}
\end{figure}

We originally chose 47 Tuc and SMC stars that did not have the same G-mag as shown in Figure~\ref{fig:47cmd2} and then corrected for the difference in magnitude with the slope of the parallax against G-mag. 

However, when parallax vs G mag in Figure \ref{fig:g_run} was fit to a line, the slopes of the three fits did not agree with each other. We then used the slope from the quasar line of best fit of 0.0054 mas/G-mag and estimated a larger error in slope of 0.0025 mas/G-mag when accounting for the parallax-magnitude systematic. Then by subtracting the average G-mag from the foreground cluster and the SMC and multiplying by the slope of 0.0054 mas/G-mag, we found the adjusted results for the distances to 47 Tuc and NGC 362 in tables \ref{tab:47summaryadj} and \ref{tab:362summaryadj}.

\begin{table}[h!]
\renewcommand{\thetable}{\arabic{table}}
\centering
\caption{Summary of results for 47 Tucanae where 47 Tuc and SMC selections have different G-mag.}
\label{tab:47summaryadj}
\begin{tabular}{cCcCC}
\tablewidth{0pt}
\hline
\hline
Method &  $\pi_{47}-\pi_{smc}$  (mas) & adjusted $\pi_{47}-\pi_{smc} (mas) $ & $\pi_{47}$ (mas) &  $d_{47} (kpc)$ \\ 
\hline
\decimals
Weighted Mean & 0.1952 \pm 0.0031 & $ 0.2067\pm0.0031\pm0.0090$ & 0.2257\pm0.0031\pm0.0108 & 4.43\pm0.06\pm0.21 \\ 
Pairwise         &0.1908 \pm 0.020& $0.2023\pm0.0020\pm0.0090$ & 0.2213\pm0.0020\pm0.0108  & 4.52\pm0.04\pm0.22 \\
Squares          &0.1982 \pm 0.0026 & $0.2097\pm0.0026\pm0.0090$ & 0.2287\pm0.0026\pm0.0108  & 4.37\pm0.05\pm0.21 \\ 
\hline
\end{tabular}
\end{table}

\begin{table}[ht!]
\renewcommand{\thetable}{\arabic{table}}
\centering
\caption{Summary of results for NGC 362 where NGC 362 and SMC selections have different G-mag.}
\label{tab:362summaryadj}
\begin{tabular}{ccccc}
\tablewidth{0pt}
\hline
\hline
Method &  $\pi_{362}-\pi_{smc} (mas)$ & adjusted $\pi_{362}-\pi_{smc} (mas)$ & $\pi_{362}$ (mas) &  $d_{362} (kpc)$ \\ 
\hline
\decimals
Weighted Mean &$0.0932\pm0.0086$& $0.1012\pm0.0086\pm0.0082$ & $0.1202\pm0.0086\pm0.0102$  & $8.32\pm0.59\pm0.70$ \\ 
Pairwise         &$0.0920\pm0.0046$& $0.1000\pm0.0046\pm0.0082$ & $0.1190\pm0.0046\pm0.0102$  & $8.40\pm0.32\pm0.72$ \\
Squares          &$0.0943\pm0.0270$& $0.1023\pm0.0270\pm0.0082$ & $0.1213\pm0.0270\pm0.0102$  & $8.2\pm1.8\pm0.7$   \\

\hline
\end{tabular}
\end{table}

The magnitude-parallax adjusted results then agree with the results previously found when using selections with the same G-mag. Due to the uncertainties introduced from correcting for the difference in apparent magnitudes, using selections of the same G-mag results in lower systematic errors and thus a more precise result. Additionally, using selections with the same G-mag results in the three methods being in better agreement for $\pi_{47}-\pi_{smc}$, suggesting that  using star selections of different G-mag can lead to inconsistencies between methods. However, selecting stars of the same apparent magnitude may not always be a possibility, thus it is important to correct for the magnitude-parallax systematic in such a case.

\end{document}